\documentstyle[12pt]{article}

\def\beqn{\begin{eqnarray}}
\def\eeqn{\end{eqnarray}}

\def\beq{\begin{equation}}
\def\eeq{\end{equation}}

\begin{document}

\begin{center}
\begin{flushright}
{\em TO THE MEMORY OF K. SELIVANOV}
\end{flushright}
\vskip 0.4cm
\begin{flushright}
{\tiny CERN-PH-TH/2004-071 \\ITEP/TH-18/04 \\{\vspace{-.2cm}}
FTPI-MINN-04/15, UMN-TH-2304/04}
\end{flushright}

\vspace{1.3cm}

\begin{center}
{\Large \bf An Exact Relation for ${\cal N}=1$ Orientifold
Field Theories with  Arbitrary Superpotential }
\end{center}

\vspace{0.3cm}

\centerline{\large  A. Armoni ${}^a$, A. Gorsky ${}^b$, M. Shifman ${}^c$}

\vskip 0.5cm
\centerline{${}^a$ {\em Department of Physics, Theory Division}}
\centerline{{\em CERN, CH-1211 Geneva 23, Switzerland}}
\vskip 0.2cm
\centerline{${}^b$ {\em Institute of
Theoretical and Experimental Physics, Moscow 177259, Russia}}
\vskip 0.2cm
\centerline{${}^c$ {\em William I. Fine Theoretical Physics Institute,
University
of Minnesota},}
\centerline{{\em Minneapolis, MN 55455, USA}}

\vskip 2cm

\begin{abstract}
We discuss a nonperturbative relation for
orientifold parent/daugh\-ter pairs of supersymmetric theories
with an arbitrary tree-level superpotential. We show that
super-Yang-Mills (SYM) theory with matter in the
adjoint representation at $N \rightarrow
\infty$, is equivalent  to a SYM theory with matter in the antisymmetric representation
and a related superpotential. The gauge
symmetry breaking patterns match in these theories too. The moduli
spaces in the limiting case of a vanishing superpotential are also discussed.
Finally we argue that there is an exact mapping between the effective
superpotentials of two finite-$N$ theories belonging to an orientifold pair.

\end{abstract}

\end{center}

\newpage

\section{Introduction}
\label{intro}

Many people hope to extract lessons from supersymmetric
(SUSY) theories that could teach us  about  nonperturbative behavior of
theories with less SUSY, hopefully, without
SUSY at all. A considerable progress was achieved in this direction.
It was discovered that theories related by orbifold or orientifold
projections are perturbatively planar equivalent
\cite{ks,lnv,bj,kakushadze}. Extension of the perturbative equivalence
to the nonperturbative level was elaborated in
\cite{Strassler:2001fs,Armoni:2003gp}. While orbifold theories
generally speaking do not enjoy nonperturbative
planar equivalence \cite{gs,Tong:2002vp}, orientifold theories do
\cite{Armoni:2003gp}. Planar equivalence means that two theories from
a given ``parent-daughter" pair have identical  behavior  at
large $N$ in common sectors. The definition of the ``common sector"
is given in the original paper \cite{Armoni:2003gp}
or in the review paper \cite{asv}.

In this letter we focus on a specific example of an
orientifold parent-daughter {\em supersymmetric} pair.
Starting from the parent softly broken ${\cal N}=2$ Yang-Mills
theory with matter in  the adjoint representation,
and a generic superpotential,
we compare it at $N \rightarrow \infty$
with a daughter super-Yang-Mills (SYM) theory with
matter in the two-index antisymmetric representation. For quadratic
superpotentials such a comparison was carried out in \cite{asv},
with the conclusion that nonperturbative planar equivalence
does take place in the common  sector of the
both theories.  The common sector includes such
non-holomorphic data as, say, the mass spectra.
Here we will consider {\em arbitrary}  superpotentials
but compare {\em only chiral data}.

We will argue that, if classical
superpotentials in both theories coincide,\footnote{The statement of
coincidence is
rather sloppy.  The required relationship between the superpotentials
in the orientifold pair is described in more accurate terms in Sect.~\ref{eesp}.}
then these theories are
planar equivalent at the nonperturbative level in the chiral sector.

More concretely, we will show that the {\em effective}
superpotentials in these theories coincide,  implying
that their holomorphic sectors  are equivalent. Furthermore,
we will demonstrate that the gauge symmetry breaking
patterns coincide in these  two theories too.
The choice of the theories above is dictated by the ability to treat them
exactly using   approaches based on  the  matrix model
\cite{Dijkgraaf:2002fc}
or the generalized Konishi anomalies in the
holomorphic sector \cite{Gorsky:2002uk,CDSW}.

First, we  demonstrate that the effective superpotentials coincide
in the two theories. To this end we use equivalence \cite{Dijkgraaf:2002xd}
of the matrix model expansion on the one hand, and  calculation
of field-theory loops in an effective background on the other hand.
This is sufficient to prove the equality of the
effective superpotentials in the planar limit.

To treat the  symmetry breaking pattern we  consider loop equations
which can be derived either from the matrix model or from
the generalized Konishi anomalies. Analysis of the loop equations
provides us with the proper identifications of the field theory
resolvents which amounts to establishing   the symmetry breaking
pattern. We exploit some results concerning antisymmetric matter
discussed  previously in
Refs.~\cite{Naculich:2003cz,Cachazo:2003kx,Kraus:2003jv,
Alday:2003gb,Landsteiner:2003rh}.

In addition, we will discuss the case of a vanishing tree-level
superpotential, namely,  correspondence between the large $N$
moduli space of ${\cal N}=2$ SYM theory and its orientifold daughter.

Finally and most importantly, we will consider a relation between
the parent and the
daughter theories at {\em finite $N$}. We will show that there exists
an exact mapping between the effective superpotential of the two
theories.

\section{Planar equivalence and orientifold field theories}
\label{pe}

The idea of planar equivalence was introduced in Ref.~\cite{ks}. It states
that two {\em distinct} gauge theories coincide at large $N$
in a certain sector. 
The original implementation of this idea was in the context of orbifolding.
While at the perturbative level the planar equivalence
does hold for orbifold field theories,
it is not valid at the non-perturbative
level \cite{gs,Tong:2002vp}.

The status of the orientifold field theories is different. In this
case both perturbative and nonperturbative proofs of
equivalence  exist; they were  given in Ref. \cite{Armoni:2003gp}.
In this section we summarize  main points of the proof. The reader
can find a more detailed discussion 
 in \cite{Armoni:2003gp} and, especially, in the review paper
\cite{asv}.

The prime example of an orientifold pair is a U($N)$ gauge theory with matter
in the adjoint representation and a U($N)$ gauge theory with matter in
the two-index antisymmetric representation (in the present paper we
consider the supersymmetric version of the former orientifold pair).

In the 't Hooft double index notation, the adjoint representation is
presented
by two lines with arrows  pointing in the opposite directions, whereas
for
the antisymmetric representation the arrows on the two lines point in
the same direction. In \cite{Armoni:2003gp} it was shown that for planar graphs
it is possible to flip the orientation of one of the arrows carrying color
flow in the
matter loops, without changing the value of any planar Feynman. This
is a perturbative proof of the planar equivalence between the two
theories.

In the same paper \cite{Armoni:2003gp} a non-perturbative proof was given as well. The main idea is that the partition
functions of the two distinct theories coincide at large $N$
{\em before} integration over the gluon field. To
this end it was demonstrated that the  determinants
in  the two theories become identical at large $N$,
\beq
 \lim _{N\rightarrow \infty}\,\,
 \frac{\det (i\not\! \partial +\not\!\!  A ^a \, T^a _{adj})}
{\det (i\not\! \partial + \not\!\!  A ^a \, T^a _{anti})}
 = 1 \, .
\eeq
This is  sufficient  to establish the nonperturbative equivalence
between two theories.

In the present paper we focus on a supersymmetric pair,
with a nonvanishing superpotential. We  demonstrate
the equivalence of the two SUSY theories
in the chiral sector by showing
that the {\em effective} superpotentials in the two theories coincide at
large $N$.

\section{Equivalence of the effective superpotentials}
\label{eesp}

To begin with, let us define the  Lagrangian of the parent SU$(N)$
theory with adjoint matter,\footnote{The subscript $F$ means that the
trace is taken in the fundamental representation, ${\rm Tr}_{F}\, W^2 \equiv
(1/2)
W^a W^a$ while $W$ is defined in such a way that
$W_\alpha = i\lambda_\alpha +...$.}
\beqn
{\cal L}_P
&=&
\frac{1}{2 g^2} \int \, d^2\theta\,  {\rm Tr}_{F}\, W^2 +
{\rm h.c.}
\nonumber\\[3mm]
&+&
\int d^2\theta d^2 \bar{\theta}
\, {\rm Tr}_{Ad} \bar{\Phi}e^{V}\Phi +
\left(
\int \, d^2 \, \theta\,  {\cal W}_{P}(\Phi)
+{\rm h.c.}\right)\,,
\eeqn
where $\Phi$ is a chiral superfield in
the adjoint and ${\cal W}_P$
is a  superpotential which will be assumed  even
in $\Phi$.

The Lagrangian of the daughter orientifold theory is
\beqn
{\cal{L}}_D
&=&
\frac{1}{2 g^2}\, \int\,  d^2\theta \,
{\rm Tr}_{F}\, W^2 +{\rm h.c.}
\nonumber\\[3mm]
&+&
\int\, d^2\theta\,  d^2 \bar{\theta} {\rm Tr}_{Anti}\left( \bar{\chi}e^{V}\chi +
\bar{\eta}e^{V}\eta\right) +\left(
\int \, d^2 \theta\,  {\cal W}_{D}(\chi\eta)
+{\rm h.c.}\right) ,
\eeqn
where $\chi,\eta$ are  antisymmetric chiral superfields
of the type $\chi^{[ij]}$, $\eta_{[ij]}$. We assume   the
tree-level superpotentials to have similar structure, namely,
\beq
{\cal W}_{P}(\Phi)=\sum_{k=1}^{k_*} g_k\, {\rm Tr}_{Ad}\, (\Phi^2)^k \,,\qquad
{\cal W}_{D}(\chi\eta)=\sum_{k=1}^{k_*} g_k \, {\rm Tr}_{Anti}\, (\chi\eta)^k\,,
\eeq
with the same coefficients $g_k$, $\, k=1,2...,k_{*}$ where
$k_{*}$ does not grow with $N$.

Turn now to the effective superpotentials in both theories.
The simplest way to derive the  effective
superpotentials is to follow Ref.~\cite{Dijkgraaf:2002xd} where it was shown
that the effective superpotentials can be calculated perturbatively
from the following effective actions
\beqn
&&\int dx d^2\theta \, \left[ \bar{\Phi}(\Delta  -iW_{\alpha}D_{\alpha})
\Phi + {\cal W}_{P}(\Phi)\right] \,, \\[3mm]
&& \int dx d^2\theta \left[ \bar{\chi}(\Delta  -iW_{\alpha}D_{\alpha})
\chi +
\bar{\eta}(\Delta  -iW_{\alpha}D_{\alpha})
\eta +
{\cal W}_{D}(\chi,\eta) \right] \,.
\eeqn
Here the chiral superfield
$W_{\alpha}$ (the gauge field strength tensor)
must be treated as a fixed constant background with
$$S=\frac{1}{32 \pi^2} {\rm Tr}_{F}W^2\,. $$
In the planar limit all graphs determining
${\cal W}_{\rm eff}(S,g_k)$ are the same, with   reversion of  the
color flow direction on one of two lines forming the loop.
Correspondingly, the result of their calculation is the
same in the parent and daughter theories at $N \rightarrow \infty$, much in
the same way as in Ref.~\cite{Armoni:2003gp}. Actually, similar arguments
with no reference to the orientifold pair were discussed previously  in
\cite{Naculich:2003cz}.

\section{The equivalence of the symmetry breaking patterns}
\label{teotsbp}

Having established the equivalence of the effective superpotentials
we pass to the symmetry breaking pattern. To this end
let us invoke another approach based on the generalized Konishi anomalies.
It was shown that they
play a crucial role in derivation of the Riemann surface which
governs the chiral sector of the theory and is equivalent to the
set of loop equations in the  matrix model.

The generalized Konishi anomalies follow from  variation
of the chiral field with a function
$f(\Phi,W)$,
$$
\Phi\rightarrow e^{f}\Phi\,.
$$
In this way one gets
\beq
\bar{D}^2 J_{f}= {\rm Tr}\left[  f(\Phi,W)\frac{\partial{\cal
W}(\Phi)}{\partial\Phi}
\right]
+ \sum_{ijkl} A_{ijkl} \frac{\partial f(\Phi,W)_{ij}}{\partial \Phi_{kl}}
\eeq
where
\beq
J_{f}={\rm Tr}\, \bar{\Phi}e^{adV}f(\Phi,W)\,,
\eeq
and the tensor $A_{ijkl}$ depends on the representation of the matter field
as follows:
\beq
A_{ijkl}= \frac{1}{32\pi^2}[W_{\alpha},[W^{\alpha},T_{lk}]]_{ji}\,.
\eeq
Here the generators of the gauge group $T_{lk}$ are taken in the
corresponding representations. It
is convenient to introduce the resolvents
\beq
T(z)= {\rm Tr}_{Ad}\, \frac{1}{z-\Phi},\qquad R(z)=
{\rm Tr}_{Ad}\,\frac{{W^2}}{z-\Phi}\,.
\eeq
Taking the function $f$ in the generalized current first to be
$T(z)$ and then   $R(z)$ one derives the following chiral ring relations:
\beqn
R^2(z)&= &{\cal W}'R(z) +\frac{1}{4}\,  q(z)\,, \nonumber\\[3mm]
2R(z)T(z) &= & {\cal W}'(z)T(z) + p(z)\,,
\eeqn
where $q(z),\,\, p(z)$ are polynomials of degree $(n-1)$
if ${\cal W}'(z)$ is a  polynomial of degree  $n$.

To get the symmetry breaking pattern one has to
calculate the integrals over the resolvent $T(z)$ over
the $A_i$ cycles over the Riemann surface defined by
the chiral ring,
\beq
\oint_{A_i} T(z) =N_i\,.
\eeq
Once $N_i$ are found one can say that the gauge group is broken as
\beq
{\rm SU}(N)\rightarrow \prod_{i}\, {\rm SU} (N_i)\,.
\eeq

Hence, to compare the symmetry breaking patterns we have to compare
the Riemann surfaces as well as the resolvents in the parent/daughter pair.
The chiral ring relation for the SU$(N)$ theory with the antisymmetric
matter has to be found from the generalized Konishi anomaly or the
matrix model. The calculation of the
generalized anomaly \cite{Naculich:2003cz} amounts to the  spectral
curve in the daughter theory, which coincides with
the one for the parent theory in the planar limit. Note that
is it convenient for our purpose to use
slightly unconventional resolvents,
\beqn
R_{D}(z) &=&
{\rm Tr}_{Anti}\,\frac{{zW^2}}{z^2-\chi \eta}
\, , \\[3mm]
T_{D}(z)&=&
{\rm Tr}_{Anti}\,\frac{z}{z^2-\chi \eta}\,.
\eeqn
Such resolvents respect the
nonanomalous U(1)$_{V}$ in the daughter theory, which results
in the fact that only the products $\chi \eta$ could develop
vacuum expectation values.
The equations for these resolvents derived from
the generalized Konishi anomalies are \cite{Argurio:2004vu}
\beqn
R^2_{D}(z)&= &{\cal W}'R_{D}(z) +\frac{1}{4}\,  q(z)\,, \\[4mm]
2R_{D}(z)T_{D}(z) &= & {\cal W}'(z)T_{D}(z) +\frac{2}{z} R_{D}(z)- 2R_{D}'(z)+ p(z)\,.
\eeqn
Comparison with similar equations in the parent theory
immediately indicates that the equations for $R(z)$ and $R_{D}(z)$
exactly coincide, while that for $T_{D}(z)$ has additional
terms compared to the parent equation. However,  these additional
terms are subleading in $N$; hence the respective equations
in the two theories
match  at large $N$, which means that the symmetry
breaking patterns coincide in the planar limit.

The example of the orientifold pair considered above is not
unique. There are several other examples of pairs with
equivalent perturbative behavior \cite{kakushadze}.
Geometrically they emerge from different orientations
of the orientifold planes in the brane picture.

Let us comment on their nonperturbative equivalence.
Consider an example of SU$(N)$ parent theory with adjoint matter and
Sp$(2N)$ daughter theory with matter in the antisymmetric representation,
or SO$(N)$ daughter with matter in the symmetric representation.

Let us exploit the exact duality found for  SU$(N)$
theory with the adjoint matter and Sp$(2N)$ theory with the
antisymmetric
matter \cite{Cachazo:2003kx} (or, alternatively,  SO$(N)$ theory
with matter in the symmetric representation \cite{Landsteiner:2003ph,Intriligator:2003xs}).
The duality implies that the SU$(N)$ theory with the adjoint matter
and superpotential ${\cal W}$ is nonperturbatively equivalent
to  Sp$(2N)$ theory provided  the following
relation between effective superpotentials takes place
\beq
{\cal W}^{\, {\rm Sp}(2N)}=\frac{1}{2}\,  {\cal W}^{\, {\rm U} (N+2k_*)} +{\rm const},
\eeq
where $k_*$ is the degree of the classical superpotentials. The
breaking patterns also match in these  two theories,
\beqn
&&
{\rm Sp}(2N)\rightarrow \prod_{k}\,{\rm Sp}(2N_k)\,,\nonumber\\[4mm]
&&
 {\rm U}(N+2k_*)\rightarrow \prod_{k}\, {\rm U}(N_k+2)\,.
\eeqn

At large $N$ we can disregard small subleading factors if
the degree of the classical superpotential $k_*$ is restricted.
The factor 1/2 is related to the fact that Ref.~\cite{Cachazo:2003kx}
deals with  only one matter field
in the antisymmetric representation, without the conjugated one;
hence,
we have to double it to get total answer for the effective superpotential.
Only in this case we have equal number of degrees of freedom in two
theories and can discuss the orientifold pair. This concludes the argument
that
nonperturbative equivalence holds for this orientifold pair as well.

\section{The moduli space in the case of a vanishing superpotential}
\label{tmscvs}

The case of a vanishing tree-level superpotential is an interesting
limiting case. The theory with adjoint matter becomes ${\cal N}=2$
super-Yang-Mills theory. It admits a classical as well as a quantum
moduli space. The exact metric on the moduli space was computed in the
case of  SU(2)  by Seiberg and Witten \cite{Seiberg:1994rs};  later
the analysis was generalized to arbitrary  SU$(N)$ in
Refs.~\cite{Klemm:1994qs,Argyres:1994xh}.

If planar equivalence holds in the limit of a vanishing superpotential
too, the orientifold daughter
should admit the same large-$N$ moduli space as the Seiberg-Witten
${\cal N}=2$ theory. While this cannot be the case literally,
a refined version of this statement is indeed
valid.

The case of a vanishing superpotential is very subtle, as all fields
are massless. Note that in the proof of non-perturbative equivalence
a small mass $m$ was needed as an infrared
regulator \cite{Armoni:2003gp}. It was assumed
that the limit $m \rightarrow 0$ is smooth. In the present situation
this limit need not be smooth, generally speaking.
In the presence of a
mass term there is no moduli space, whereas the theories develop a moduli
space when $m=0$. In order to demonstrate the subtlety of the limit $m
\rightarrow 0$, we quote a result  \cite{asv} for the ratio of matter
condensates in the two theories at finite $N$,
\beq
\frac{m_{D}\, \langle\xi\eta\rangle}{m_{P}\, \langle\Phi^2\rangle}
=\frac{N-2}{N}\, (8\pi ^2)^{\frac{4}{N^2}-\frac{4}{N}}\left (
1+\frac{1}{N} \right )^
\frac{4(N-1)}{N^2} \,
\left(\frac{\Lambda}{m}\right)^{2/N}
\, .
\eeq
It is clear that as the ratio $m/\Lambda$ decreases,
a critical value of $N$ needed for the onset
of the planar equivalence increases logarithmically,
\beq
N_*\sim \ln \frac{\Lambda}{m}\,.
\eeq
Thus, our analysis in the following assumes the strict planar limit, $N=\infty$.
The above example demonstrates that the limit $m\rightarrow 0$ and
$1/N \rightarrow 0$ might not commute.

We turn now to the check of planar equivalence. The large-$N$
limit  in the case
of a moduli space means that the matter vacuum expectation values
should scale as $\sqrt N$, as we want to keep the $W$
masses fixed (recall the the $W$ mass is $M_W = g v$).

Let us analyze first the classical moduli space of the two theories.
Let us assume, for simplicity that $N$ is even (the analysis in the
odd-$N$  case is straightforward too).
The classical moduli of ${\cal N}=2$ are
\beq
u_k = \langle\, {\rm Tr}_{Ad}\, \Phi ^k\, \rangle \,,\qquad  k=1,..., N \, .
\eeq
Since ${\rm Tr}\, \Phi =0$ there are actually only $N-1$ moduli.
The classical moduli space of the daughter theory with the
antisymmetric matter is
\beq
v_k = \langle\,{\rm Tr}_{Anti} (\chi \eta)^k \, \rangle  \,,\qquad  k=1,..., N/2 \, .
\eeq
Clearly the two moduli spaces do not match, even at the classical
level.

Note, however, that the correspondence that we suggest is between
the operators $ (\chi \eta)^k$ and $ (\Phi ^2)^k $. In order to compare
the two moduli spaces we should restrict ourselves to the subspace
of ${\cal N}=2$ SYM
theory where the odd moduli are frozen, ${\rm Tr}_{Ad} \, \Phi ^{2k+1}
=0$.
Thus, we suggest a correspondence (or, actually, an equivalence) between the
large-$N$ classical moduli
space of the orientifold daughter and the even moduli subspace of
${\cal N}=2$ SYM theory.

Let us discuss now the quantum moduli space. The Seiberg-Witten curve
of the ${\cal N}=2$ theory is
\beq
y^2 + y(x^N + u_2 x^{N-2} + u_3 x^{N-3} + ... + u_N ) + 1 =0 \, .
\eeq
Another way of writing the Seiberg-Witten curve, that has a nice
interpretation in terms of locations of D4 branes at $x=\{a_i \}$, is
\cite{Witten:1997sc}
\beq
y^2 + y(x-a_1)(x-a_2)...(x-a_N) + 1 =0 \, .
\eeq
The relation between the different parameterizations of the moduli, $\{
a_i \}$ and $\{ u_i \}$, is obtained by comparing the coefficients in
front of $x^i$. In particular, $\sum a_i =0$.
Restricting ourselves to the case where the odd moduli are frozen,
$$u_{2k+1} =0\,,$$ the Seiberg-Witten curve of this subspace can be
written in terms of $N/2$ moduli $\{ b_i \} $ as follows:
\beq
y^2 + y(x-b_1)(x+b_1)(x-b_2)(x+b_2)...(x-b_{N/2})(x+b_{N/2}) + 1 =0 \, ,
\eeq
or
\beq
y^2 + y(x^2-b_1^2)(x^2-b_2^2)...(x^2-b_{N/2}^2) + 1 =0 \, .
\eeq
This is just the expected Seiberg-Witten curve for the SU$(N)$ orientifold theory
at large $N$.

It was obtained previously by brane techniques in
Refs.~\cite{Csaki:1998mx,Park:1998kd,Landsteiner:1998pb}.
The above result clearly reflects the presence of an orientifold plane in
the brane picture: the branes are located on both sides of the
orientifold at the following positions:
$$x=b_1, \,\, x=-b_1, \,\, x=b_2, \,\, x=-b_2, ... ,\,\,
x=b_{N/2}, \,\, x=-b_{N/2}\,.$$
This is also the solution for SO$(N)$ theories at large $N$.

Note that the equivalence of the Seiberg-Witten curves, at large-$N$,
implies a coincidence of the BPS spectra of the two theories (up to
$1/N$
corrections). A similar result for a SUSY/non-SUSY orientifold pair was already
obtained in \cite{Armoni:2003gp}.

Let us emphasize that matching of the moduli space generically is true
only for large $N$;  there is an evident counterexample for $N=3$. Indeed,
in the parent SU(3)  ${\cal N}=2$ theory there is a two-dimensional
complex moduli space.
At the same time, let us examine possible moduli space in the
daughter SU(3) ${\cal N}=1$ theory with $N_f=1$. First, note that there
is no Higgs branch of the moduli space, which implies the $N_f>1$ condition.
On the other hand, it is known that there is no Coulomb branch of the moduli
space in this theory either; hence, we have clear mismatch between
the parent and daughter theories at $N=3$.

Though we are interested only in the large $N$ limit, we wish to
comment on the odd $N$ case. Here we have $(N-1)/2$ moduli
characterized by
\beq
v_k = \langle\,{\rm Tr}_{Anti} (\chi \eta)^k \, \rangle \,,\qquad  k=1, ..., (N-1)/2 \, .
\eeq
It should be compared with the ${\cal N}=2$ theory with odd $N$,
where the odd moduli are frozen, $u_{2i+1}=0,\,\,\, i=1, ..., (N-1)/2$.
The Seiberg-Witten curve in this case is
\beq
y^2 + yx(x^2-b_1^2)(x^2-b_2^2)...(x^2-b_{(N-1)/2}^2) + 1 =0 \, .
\eeq

Let us briefly compare our case with the {\em orbifold} (rather than orientifold)
daughter. In the
orbifold pair the  ${\cal N}=2$ parent theory gets mapped onto
${\cal N}=1$ daughter with bifundamental matter. In the orbifold
case   comparison of the Seiberg-Witten curves of the pair
was performed in \cite{erlich} where it was argued that
in the large-$N$ limit the curves match if  all moduli in the parent
theory which are not singlets under the orbifold group are set to zero.
This is similar to the orientifold case under consideration.
However, there is a difference between the two cases in the
second ingredient of the solution to  ${\cal N}=2$ theory, namely
the differential on the curve. In the orbifold case the differential
gets changed resulting in  a rescaling of the coupling constant in
the daughter theory while in the orientifold case the coupling constants
in the parent and daughter theories are the same.

In conclusion, it is possible to construct the large-$N$
Seiberg--Witten curves of an orbifold/orientifold daughter theories
by keeping the moduli  invariant  under the orbifold/orientifold action and
projecting out the noninvariant moduli. An interesting question that
we wish to pose here is:
``can one obtain   {\em all} field theories which admit a
Seiberg-Witten curve and a brane realization in string theory
 by a suitable orbifold/orientifold projection?"

\section{Relation for the orientifold pair at finite $N$}
\label{finite}

So far we discussed the equivalence in the orientifold
pairs at large $N$. In this section we wish to make a much
stronger statement of a relation between the two theories
at finite $N$.

Let us consider first the $N$ dependence in the U$(N)$ theory
with matter in the adjoint representation. In such a theory
the effective superpotential takes the following form:
\beq
{\cal W}_{\rm eff} = N (S\log S -S) + N {\cal W}_{\rm pert}(S)\,,
\label{UN}
\eeq
where $ {\cal W}_{\rm pert}(S)$ (the ``perturbative'' part = the polynomial part of the
effective action) follows from a matrix model integral.
In particular, the whole action is proportional to $N$.
The linear $N$ dependence in the polynomial part of the
effective action ${\cal W}_{\rm pert}$
follows from the Dijkgraaf-Vafa prescription for
calculating ${\cal W}_{\rm pert}(S)$,
\beq
{\cal W}_{\rm pert}(S) = {\partial {\cal F}_0 \over \partial S}\,,
\eeq
where ${\cal F}_0 = {\cal F}_0 (S,g_k)$.

Now let us ask what is the form of the resulting effective superpotential
in the theory with the antisymmetric matter. As was argued above,
at large $N$ it must coincide with the
effective superpotential (\ref{UN}) of the U$(N)$ theory coupled to
the adjoint matter. Moreover, since for  U(2)  the antisymmetric
representation is in fact a singlet, the matter decouples
from the gauge part, and the effective superpotential must reduce to the
Veneziano-Yankielowicz action \cite{VenY}. The unique solution that meets those twin requirements is
\beq
{\cal W}_{\rm eff} = N (S\log S -S) + (N-2) {\cal W}_{\rm pert}(S)\,.
\label{antisymmetric}
\eeq
Note a crucial point: nonpolynomial factors such as $N-4/N$, that also vanish at
$N=2$, are ruled out, as they
have  no meaning in terms of the Ramond-Ramond fluxes in the brane picture.

The $N-2$ factor in front of the effective superpotential in the theories
that include orientifold planes, was already observed in
\cite{Landsteiner:2003rh,Landsteiner:2003ua}.
Thus, the knowledge of the effective superpotential in one of the
theories  automatically fixes the effective superpotential in the
other. It is interesting to note that for  SU(3), where the
antisymmetric representation is equivalent to  antifundamental, we can relate
the actions of the theory with the adjoint and fundamental matter.
This is quite remarkable!

As an example, let us consider a particular case with a tree-level 
{\em quartic}
superpo\-tential.\footnote{We thank R. Argurio for bringing this example
to our attention.}
The effective superpotential for the theory with the
adjoint matter was calculated in Ref.~\cite{Fuji:2002wd};
for the theory with the
fundamental matter it was calculated in Ref.~\cite{Argurio:2002xv}. 
For SU(3) the
antifundamental representation is equivalent to
two-index  antisymmetric. However, 
  Refs.~\cite{Fuji:2002wd,Argurio:2002xv} deal with
U$(N)$ group rather than SU$(N)$, and, therefore,
we need to  translate  their results
from U$(N)$ to SU$(N)$. This translation, together with the relations
$$
Q^i = {1\over 2} \epsilon ^{ijk} \chi _{jk}\,,\qquad \bar Q_i ={1\over
2}\epsilon _{ijk} \eta ^{[jk]}\,,
$$
results  in  a relation between the
tree-level couplings of the two theories.
This result can be expressed as follows:
the effective superpotentials in
both theories take  the same form, except for an
anticipated  overall factor (we use the notation of Ref.~\cite{Fuji:2002wd}),
\beq
{\cal W}^ {\rm adjoint}_{\rm pert}=3{\cal W}^ {\rm fundamental}_{\rm pert}=
 - \sum _{k=1} ^\infty \left (-{3g \over 2m^2}
\right )^k S^{k+1}
{(2k-1)! \over k! (k+1)!}\,\,\, .
\eeq
Let us note that the agreement between two  SU(3)  theories with  
quartic superpotentials on the one hand,
and a mismatch in moduli  in the limit  ${\cal W}_{\rm tree}=0$
on the other,
presumably implies that in the process
of switching off the superpotential in the daughter theory we
arrive at a singularity point in
the moduli space of the ${\cal N}=2$  SU(3)  parent theory.

Similar arguments can be applied to the case with general
symmetry breaking patterns. Namely, if we consider the
${\rm U}(N)\rightarrow \prod_{i}\,{\rm U}(N_i)$ pattern we have the same
twin requirements again. Indeed, we have already argued that
at large $N$ the symmetry breaking patterns in the orientifold
pair match  while in each $U(N_i)$ matter decouples if $N_i=2$.
Hence, we have a similar expression in the generic case,
\beq
{\cal W}_{\rm eff} = \sum_i (N_i (S_i\log S_i -S_i) + (N_i-2) {\cal W}_{\rm
pert}(S_i))\, .
\label{antisymmetric2}
\eeq
Finally, it is rather clear that in  the theory with the symmetric matter the
effective superpotential is
\beq
{\cal W}_{\rm eff} = N (S\log S -S) + (N+2) {\cal W}_{\rm pert}(S)\, .
\label{symmetric}
\eeq

\section{Discussion}
\label{disc}

In this paper we   show  that holomorphic data
in the orientifold pair with $\cal{N}=1$   match
in the planar limit. To an extent this follows
from the possibility of deriving the effective potential
perturbatively in the  coupling constants using the formalism of the external constant
composite background field $S$. Although these data do not cover
the whole content of the theory, the nonperturbative planar
equivalence we observe means that the domain wall tensions in two theories
coincide. One could also consider the domain wall junctions
saturating the central charge in the anticommutator of ${\bar Q}$ and $Q$.
This central charge does not belong to the holomorphic sector but it
involves the matter axial currents which can be mapped between
the two theories. Hence, it is plausible  that the tensions of
the domain wall junctions coincide
in these theories at large $N$ as well.

Finally, we wish to comment on the finite-$N$ relation that we established
between the two SUSY theories.  The relation amounts to  a simple
shift $N \rightarrow N-2$. It could be extremely useful for
phenomenology if this feature extends   to certain
quantities in the case of the SUSY/non-SUSY pair.

\vspace{0.2cm}

{\bf Acknowledgments} A. A. thanks R. Argurio, K. Landsteiner, 
E. Lopez, G. Veneziano and N. Wyllard for useful discussions.
A.G. thanks
FTPI at University of Minnesota, where this work was partially done, for
the kind hospitality. The work  of M. S. is
supported in part by DOE grant DE-FG02-94ER408.

\vspace{2mm}

{\bf Note added. July 21, 2004.}

After this work was completed
and submitted for publication in Nuclear Physics, 
two related papers by Argurio
and Landsteiner were posted on the electronic archive 
\cite{Landsteiner:2004ng,Argurio:2004vu}. These two works
treat  effective superpotentials in theories with antisymmetric
matter, and both works support our assertion of planar equivalence.

\end{document}